\documentclass[journal]{IEEEtran}
\usepackage{multicol}
\usepackage{graphicx}
\graphicspath{{Figures/}}
\usepackage{epstopdf}
\usepackage{acronym}
\usepackage{amsmath}
\usepackage{amsfonts}
\usepackage{color, colortbl}
\usepackage{subfigure}
\usepackage[english]{babel}
\usepackage{blindtext}
\usepackage{algorithm} 
\usepackage{algorithmic} 
\usepackage{multirow}
\usepackage{color}
\usepackage{subfigure}
\usepackage{balance}
\usepackage{caption}
\usepackage{gensymb}

\begin{document}

\newacro{3GPP}{third generation partnership project}
\newacro{4G}{4{th} generation}
\newacro{5G}{5{th} generation}

\newacro{Adam}{adaptive moment estimation}
\newacro{ADC}{analogue-to-digital converter}
\newacro{AED}{accumulated euclidean distance}
\newacro{AGC}{automatic gain control}
\newacro{AI}{artificial intelligence}
\newacro{AMB}{adaptive multi-band}
\newacro{AMB-SEFDM}{adaptive multi-band SEFDM}
\newacro{AN}{artificial noise}
\newacro{ANN}{artificial neural network}
\newacro{ASE}{amplified spontaneous emission}
\newacro{ASIC}{application specific integrated circuit}
\newacro{AWG}{arbitrary waveform generator}
\newacro{AWGN}{additive white Gaussian noise}
\newacro{A/D}{analog-to-digital}

\newacro{B2B}{back-to-back}
\newacro{BCF}{bandwidth compression factor}
\newacro{BCJR}{Bahl-Cocke-Jelinek-Raviv}
\newacro{BDM}{bit division multiplexing}
\newacro{BED}{block efficient detector}
\newacro{BER}{bit error rate}
\newacro{Block-SEFDM}{block-spectrally efficient frequency division multiplexing}
\newacro{BLER}{block error rate}
\newacro{BPSK}{binary phase shift keying}
\newacro{BS}{base station}
\newacro{BSS}{best solution selector}
\newacro{BU}{butterfly unit}

\newacro{CapEx}{capital expenditure}
\newacro{CA}{carrier aggregation}
\newacro{CBS}{central base station}
\newacro{CC}{component carriers}
\newacro{CCDF}{complementary cumulative distribution function}
\newacro{CCE}{control channel element}
\newacro{CCs}{component carriers}
\newacro{CD}{chromatic dispersion}
\newacro{CDF}{cumulative distribution function}
\newacro{CDI}{channel distortion information}
\newacro{CDMA}{code division multiple access}
\newacro{CI}{constructive interference}
\newacro{CIR}{carrier-to-interference power ratio}
\newacro{CMOS}{complementary metal-oxide-semiconductor}
\newacro{CNN}{convolutional neural network}
\newacro{CoMP}{coordinated multiple point}
\newacro{CO-SEFDM}{coherent optical-SEFDM}
\newacro{CP}{cyclic prefix}
\newacro{CPE}{common phase error}
\newacro{CRVD}{conventional real valued decomposition}
\newacro{CR}{cognitive radio}
\newacro{CRC}{cyclic redundancy check}
\newacro{CS}{central station}
\newacro{CSI}{channel state information}
\newacro{CSIT}{channel state information at transmitter}
\newacro{CSPR}{carrier to signal power ratio}
\newacro{CW}{continuous-wave}
\newacro{CWT}{continuous wavelet transform}
\newacro{C-RAN}{cloud-radio access networks}

\newacro{DAC}{digital-to-analogue converter}
\newacro{DBP}{digital backward propagation}
\newacro{DC}{direct current}
\newacro{DCGAN}{deep convolutional generative adversarial network}
\newacro{DCI}{downlink control information}
\newacro{DCT}{discrete cosine transform}
\newacro{DDC}{digital down-conversion}
\newacro{DDO-OFDM}{directed detection optical-OFDM}
\newacro{DDO-OFDM}{direct detection optical-OFDM}
\newacro{DDO-SEFDM}{directed detection optical-SEFDM}
\newacro{DFB}{distributed feedback}
\newacro{DFDMA}{distributed FDMA}
\newacro{DFT}{discrete Fourier transform}
\newacro{DFrFT}{discrete fractional Fourier transform}
\newacro{DL}{deep learning}
\newacro{DMA}{direct memory access}
\newacro{DMRS}{demodulation reference signal}
\newacro{DoF}{degree of freedom}
\newacro{DOFDM}{dense orthogonal frequency division multiplexing}
\newacro{DP}{dual polarization}
\newacro{DPC}{dirty paper coding}
\newacro{DSB}{double sideband}
\newacro{DSL}{digital subscriber line}
\newacro{DSP}{digital signal processors}
\newacro{DSSS}{direct sequence spread spectrum}
\newacro{DT}{decision tree}
\newacro{DVB}{digital video broadcast}
\newacro{DWDM}{dense wavelength division multiplexing}
\newacro{DWT}{discrete wavelet transform}
\newacro{D/A}{digital-to-analog}

\newacro{ECC}{error correcting codes}
\newacro{ECL}{external-cavity laser}
\newacro{ECOC}{error-correcting output codes}
\newacro{EDFA}{erbium doped fiber amplifier}
\newacro{EE}{energy efficiency}
\newacro{eMBB}{enhanced mobile broadband}
\newacro{eNB-IoT}{enhanced NB-IoT}
\newacro{EPA}{extended pedestrian A}
\newacro{EVM}{error vector magnitude}

\newacro{Fast-OFDM}{fast-orthogonal frequency division multiplexing}
\newacro{FBMC}{filterbank based multicarrier }
\newacro{FCE}{full channel estimation}
\newacro{FD}{fixed detector}
\newacro{FDD}{frequency division duplexing}
\newacro{FDM}{frequency division multiplexing}
\newacro{FDMA}{frequency division multiple access}
\newacro{FE}{full expansion}
\newacro{FEC}{forward error correction}
\newacro{FEXT}{far-end crosstalk}
\newacro{FF}{flip-flop}
\newacro{FFT}{fast Fourier transform}
\newacro{FFTW}{Fastest Fourier Transform in the West}
\newacro{FHSS}{frequency-hopping spread spectrum}
\newacro{FIFO}{first in first out}
\newacro{FMCW}{frequency-modulated continuous wave}
\newacro{F-OFDM}{filtered-orthogonal frequency division multiplexing}
\newacro{FPGA}{field programmable gate array}
\newacro{FrFT}{fractional Fourier transform}
\newacro{FSD}{fixed sphere decoding}
\newacro{FSD-MNSF}{FSD-modified-non-sort-free}
\newacro{FSD-NSF}{FSD-non-sort-free}
\newacro{FSD-SF}{FSD-sort-free}
\newacro{FSK}{frequency shift keying}
\newacro{FTN}{faster than Nyquist}
\newacro{FTTB}{fiber to the building}
\newacro{FTTC}{fiber to the cabinet}
\newacro{FTTdp}{fiber to the distribution point}
\newacro{FTTH}{fiber to the home}

\newacro{GAN}{generative adversarial network}
\newacro{GB}{guard band}
\newacro{GFDM}{generalized frequency division multiplexing}
\newacro{GPU}{graphics processing unit}
\newacro{GSM}{global system for mobile communication}
\newacro{GUI}{graphical user interface}

\newacro{HARQ}{hybrid automatic repeat request}
\newacro{HC-MCM}{high compaction multi-carrier communication}
\newacro{HPA}{high power amplifier}

\newacro{IC}{integrated circuit}
\newacro{ICI}{inter carrier interference}
\newacro{ID}{iterative detection}
\newacro{IDCT}{inverse discrete cosine transform}
\newacro{IDFT}{inverse discrete Fourier transform}
\newacro{IDFrFT}{inverse discrete fractional Fourier transform}
\newacro{ID-FSD}{iterative detection-FSD}
\newacro{ID-SD}{ID-sphere decoding}
\newacro{IF}{intermediate frequency}
\newacro{IFFT}{inverse fast Fourier transform}
\newacro{IFrFT}{inverse fractional Fourier transform}
\newacro{IM}{index modulation}
\newacro{IMD}{intermodulation distortion}
\newacro{IoT}{internet of things}
\newacro{IOTA}{isotropic orthogonal transform algorithm}
\newacro{IP}{intellectual property}
\newacro{IR}{infrared}
\newacro{ISAC}{integrated sensing and communication}
\newacro{ISAR}{inverse synthetic aperture radar}
\newacro{ISC}{interference self cancellation}
\newacro{ISI}{inter symbol interference}
\newacro{ISM}{industrial, scientific and medical}
\newacro{ISTA}{iterative shrinkage and thresholding}

\newacro{KNN}{k-nearest neighbours}

\newacro{LDPC}{low density parity check}
\newacro{LFDMA}{localized FDMA}
\newacro{LLR}{log-likelihood ratio}
\newacro{LNA}{low noise amplifier}
\newacro{LO}{local oscillator}
\newacro{LOS}{line-of-sight}
\newacro{LPWAN}{low power wide area network}
\newacro{LS}{least square}
\newacro{LSTM}{long short-term memory}
\newacro{LTE}{long term evolution}
\newacro{LTE-Advanced}{long term evolution-advanced}
\newacro{LUT}{look-up table}

\newacro{MA}{multiple access}
\newacro{MAC}{media access control}
\newacro{MAMB}{mixed adaptive multi-band}
\newacro{MAMB-SEFDM}{mixed adaptive multi-band SEFDM}
\newacro{MASK}{m-ary amplitude shift keying}
\newacro{MB}{multi-band}
\newacro{MB-SEFDM}{multi-band SEFDM}
\newacro{MCM}{multi-carrier modulation}
\newacro{MC-CDMA}{multi-carrier code division multiple access}
\newacro{MCS}{modulation and coding scheme}
\newacro{MF}{matched filter}
\newacro{MIMO}{multiple input multiple output}
\newacro{ML}{maximum likelihood}
\newacro{MLSD}{maximum likelihood sequence detection}
\newacro{MMF}{multi-mode fiber}
\newacro{MMSE}{minimum mean squared error}
\newacro{mMTC}{massive machine-type communication}
\newacro{MNSF}{modified-non-sort-free}
\newacro{MOFDM}{masked-OFDM}
\newacro{MRVD}{modified real valued decomposition}
\newacro{MS}{mobile station}
\newacro{MSE}{mean squared error}
\newacro{MTC}{machine-type communication}
\newacro{MUI}{multi-user interference}
\newacro{MUSA}{multi-user shared access}
\newacro{MU-MIMO}{multi-user multiple-input multiple-output}
\newacro{MZM}{Mach-Zehnder modulator}
\newacro{M2M}{machine to machine}

\newacro{NB-IoT}{narrowband IoT}
\newacro{NB}{naive Bayesian}
\newacro{NDFF}{National Dark Fiber Facility}
\newacro{NEXT}{near-end crosstalk}
\newacro{NFV}{network function virtualization}
\newacro{NG-IoT}{next generation IoT}
\newacro{NLOS}{non-line-of-sight}
\newacro{NLSE}{nonlinear Schrödinger equation}
\newacro{NN}{neural network}
\newacro{NOFDM}{non-orthogonal frequency division multiplexing}
\newacro{NOMA}{non-orthogonal multiple access}
\newacro{NoFDMA}{non-orthogonal frequency division multiple access}
\newacro{NP}{non-polynomial}
\newacro{NR}{new radio}
\newacro{NSF}{non-sort-free}
\newacro{NWDM}{Nyquist wavelength division multiplexing }
\newacro{Nyquist-SEFDM}{Nyquist-spectrally efficient frequency division multiplexing}

\newacro{OBM-OFDM}{orthogonal band multiplexed OFDM}
\newacro{OF}{optical filter}
\newacro{OFDM}{orthogonal frequency division multiplexing}
\newacro{OFDMA}{orthogonal frequency division multiple access}
\newacro{OMA}{orthogonal multiple access}
\newacro{OpEx}{operating expenditure}
\newacro{OPM}{optical performance monitoring}
\newacro{OQAM}{offset-QAM}
\newacro{OSI}{open systems interconnection}
\newacro{OSNR}{optical signal-to-noise ratio}
\newacro{OSSB}{optical single sideband}
\newacro{OTA}{over-the-air}
\newacro{OTFS}{orthogonal time frequency space}
\newacro{Ov-FDM}{Overlapped FDM}
\newacro{O-SEFDM}{optical-spectrally efficient frequency division multiplexing}
\newacro{O-FOFDM}{optical-fast orthogonal frequency division multiplexing}
\newacro{O-OFDM}{optical-orthogonal frequency division multiplexing}
\newacro{O-CDMA}{optical-code division multiple access}

\newacro{PA}{power amplifier}
\newacro{PAPR}{peak-to-average power ratio}
\newacro{PCA}{principal component analysis}
\newacro{PCE}{partial channel estimation}
\newacro{PD}{photodiode}
\newacro{PDCCH}{physical downlink control channel}
\newacro{PDF}{probability density function}
\newacro{PDP}{power delay profile}
\newacro{PDMA}{polarisation division multiple access}
\newacro{PDM-OFDM}{polarization-division multiplexing-OFDM}
\newacro{PDM-SEFDM}{polarization-division multiplexing-SEFDM}
\newacro{PDSCH}{physical downlink shared channel}
\newacro{PE}{processing element}
\newacro{PED}{partial Euclidean distance}
\newacro{PLA}{physical layer authentication}
\newacro{PLS}{physical layer security}
\newacro{PMD}{polarization mode dispersion}
\newacro{PON}{passive optical network}
\newacro{PPM}{parts per million}
\newacro{PRB}{physical resource block}
\newacro{PSD}{power spectral density}
\newacro{PSK}{pre-shared key}
\newacro{PSS}{primary synchronization signal}
\newacro{PU}{primary user}
\newacro{PXI}{PCI extensions for instrumentation}
\newacro{P/S}{parallel-to-serial}

\newacro{QAM}{quadrature amplitude modulation}
\newacro{QKD}{quantum key distribution}
\newacro{QoS}{quality of service}
\newacro{QPSK}{quadrature phase-shift keying}
\newacro{QRNG}{quantum random number generation}

\newacro{RAUs}{remote antenna units}
\newacro{RBF}{radial basis function}
\newacro{RBW}{resolution bandwidth}
\newacro{ReLU}{rectified linear units}
\newacro{RF}{radio frequency}
\newacro{RMS}{root mean square}
\newacro{RMSE}{root mean square error}
\newacro{RMSProp}{root mean square propagation}
\newacro{RNTI}{radio network temporary identifier}
\newacro{RoF}{radio-over-fiber}
\newacro{ROM}{read only memory}
\newacro{RRC}{root raised cosine}
\newacro{RSC}{recursive systematic convolutional}
\newacro{RSSI}{received signal strength indicator}
\newacro{RTL}{register transfer level}
\newacro{RVD}{real valued decomposition}

\newacro{SB-SEFDM}{single-band SEFDM}
\newacro{ScIR}{sub-carrier to interference ratio}
\newacro{SCMA}{sparse code multiple access}
\newacro{SC-FDMA}{single carrier frequency division multiple access}
\newacro{SC-SEFDMA}{single carrier spectrally efficient frequency division multiple access}
\newacro{SD}{sphere decoding}
\newacro{SDM}{space division multiplexing}
\newacro{SDMA}{space division multiple access}
\newacro{SDN}{software-defined network}
\newacro{SDP}{semidefinite programming}
\newacro{SDR}{software-defined radio}
\newacro{SE}{spectral efficiency}
\newacro{SEFDM}{spectrally efficient frequency division multiplexing}
\newacro{SEFDMA}{spectrally efficient frequency division multiple access} 
\newacro{SF}{sort-free}
\newacro{SFCW}{stepped-frequency continuous wave}
\newacro{SGD}{stochastic gradient descent}
\newacro{SGDM}{stochastic gradient descent with momentum}
\newacro{SIC}{successive interference cancellation}
\newacro{SiGe}{silicon-germanium}
\newacro{SINR}{signal-to-interference-plus-noise ratio}
\newacro{SIR}{signal-to-interference ratio}
\newacro{SISO}{single-input single-output}
\newacro{SLM}{spatial light modulator}
\newacro{SMF}{single mode fiber}
\newacro{SNR}{signal-to-noise ratio}
\newacro{SP}{shortest-path}
\newacro{SPSC}{symbol per signal class}
\newacro{SPM}{self-phase modulation}
\newacro{SRS}{sounding reference signal}
\newacro{SSB}{single-sideband}
\newacro{SSBI}{signal-signal beat interference}
\newacro{SSFM}{split-step Fourier method}
\newacro{SSMF}{standard single mode fiber}
\newacro{STBC}{space time block coding}
\newacro{STFT}{short time Fourier transform}
\newacro{STC}{space time coding}
\newacro{STO}{symbol timing offset}
\newacro{SU}{secondary user}
\newacro{SVD}{singular value decomposition}
\newacro{SVM}{support vector machine}
\newacro{SVR}{singular value reconstruction}
\newacro{S/P}{serial-to-parallel}

\newacro{TDD}{time division duplexing}
\newacro{TDMA}{time division multiple access }
\newacro{TDM}{time division multiplexing}
\newacro{TFP}{time frequency packing}
\newacro{THP}{Tomlinson-Harashima precoding}
\newacro{TOFDM}{truncated OFDM}
\newacro{TSPSC}{training symbols per signal class}
\newacro{TSVD}{truncated singular value decomposition}
\newacro{TSVD-FSD}{truncated singular value decomposition-fixed sphere decoding}
\newacro{TTI}{transmission time interval}

\newacro{UAV}{unmanned aerial vehicle}
\newacro{UCR}{user compression ratio}
\newacro{UE}{user equipment}
\newacro{UFMC}{universal-filtered multi-carrier}
\newacro{ULA}{uniform linear array}
\newacro{UMTS}{universal mobile telecommunications service}
\newacro{URLLC}{ultra-reliable low-latency communication}
\newacro{USRP}{universal software radio peripheral}
\newacro{UWB}{ultra-wideband}

\newacro{VDSL}{very-high-bit-rate digital subscriber line}
\newacro{VDSL2}{very-high-bit-rate digital subscriber line 2}
\newacro{VHDL}{very high speed integrated circuit hardware description language}
\newacro{VLC}{visible light communication}
\newacro{VLSI}{very large scale integration}
\newacro{VOA}{variable optical attenuator}
\newacro{VP}{vector perturbation}
\newacro{VSSB-OFDM}{virtual single-sideband OFDM}
\newacro{V2V}{vehicle-to-vehicle}

\newacro{WAN}{wide area network}
\newacro{WCDMA}{wideband code division multiple access}
\newacro{WDM}{wavelength division multiplexing}
\newacro{WDP}{waveform-defined privacy}
\newacro{WDS}{waveform-defined security}
\newacro{WiFi}{wireless fidelity}
\newacro{WiGig}{Wireless Gigabit Alliance}
\newacro{WiMAX}{Worldwide interoperability for Microwave Access}
\newacro{WLAN}{wireless local area network}
\newacro{WSS}{wavelength selective switch}

\newacro{XPM}{cross-phase modulation}

\newacro{ZF}{zero forcing}
\newacro{ZP}{zero padding}


\title{An Experimental Proof of Concept for Integrated Sensing and Communications Waveform Design}

\author{{Tongyang Xu,~\IEEEmembership{Member,~IEEE}, Fan Liu,~\IEEEmembership{Member,~IEEE}, Christos Masouros,~\IEEEmembership{Senior Member,~IEEE} and Izzat Darwazeh,~\IEEEmembership{Senior Member,~IEEE}}
\thanks{
T. Xu, C. Masouros and I. Darwazeh are with the Department of Electronic and Electrical Engineering, University College London (UCL), London, WC1E 7JE, UK (e-mail: tongyang.xu.11@ucl.ac.uk, c.masouros@ucl.ac.uk, i.darwazeh@ucl.ac.uk). 

Fan Liu is with the Department of Electical and Electronic Engineering, Southern University of Science and Technology, Shenzhen, China. e-mail: liuf6@sustech.edu.cn. 

This work was supported by the Engineering and Physical Sciences Research Council (EPSRC) general Grant EP/S028455/1.
}}

\maketitle

\begin{abstract}

The integration of sensing and communication (ISAC) functionalities have recently gained significant research interest as a hardware-, power-, spectrum- and cost- efficient solution. This experimental work focuses on a dual-functional radar sensing and communication framework where a single radiation waveform, either omnidirectional or directional, can realize both radar sensing and communication functions. We study a trade-off approach that can balance the performance of communications and radar sensing. We design an orthogonal frequency division multiplexing (OFDM) based multi-user multiple input multiple output (MIMO) software-defined radio (SDR) testbed to validate the dual-functional model. We carry out over-the-air experiments to investigate the optimal trade-off factor to balance the performance for both functions. On the radar performance, we measure the output beampatterns of our transmission to examine their similarity to simulation based beampatterns. On the communication side, we obtain bit error rate (BER) results from the testbed to show the communication performance using the dual-functional waveform. Our experiment reveals that the dual-functional approach can achieve comparable BER performance with pure communication-based solutions while maintaining fine radar beampatterns simultaneously. 

\end{abstract}


\section{Introduction}

\IEEEPARstart{W}{ireless} communications have evolved from 1G to 5G with significant technology innovations. Traditionally, signals are transmitted at low-frequency carriers with narrow signal bandwidth due to limitations from hardware and technical theories. Nowadays, signals can be transmitted at millimeter wave (mmWave) frequency \cite{Rappaport_book2014} and TeraHertz (THz) frequency \cite{THz_2011} with GHz signal bandwidth. In terms of antennas, communication systems can integrate hundreds of antennas in massive \ac{MIMO} \cite{MIMO_massive_Erik_2014}. Moreover, in terms of signal waveform, different options are available such as \ac{CDMA} in 3G \cite{Erik_book_3G}, \ac{OFDM} and \ac{SC-FDMA} in 4G/5G \cite{Erik_book_4G, Erik_book_5G}. Recently, advanced waveform candidates are being investigated for future 6G such as \ac{SEFDM} \cite{TongyangTVT2017}, \ac{FTN} \cite{Anderson2013}, \ac{OTFS} \cite{OTFS_WCNC2017}, \ac{GFDM} \cite{GFDM_trans} and \ac{FBMC} \cite{FBMC2011_magazine}.

Complementary to wireless communications, various sensors have been used to sense the world such as accelerometers, Gyroscope, light sensor, temperature sensors, audio and video. Due to the ubiquitous features of wireless signals, smart applications such as non-intrusive and non-contact radar sensing and \ac{RF} sensing are becoming popular. In \cite{Sensing_Soli_Google_2016}, Google develops a mmWave radar sensing system at 60 GHz termed `Soli', which can sense and understand subtle motions in finger gestures. Work in \cite{Sensing_radar_sonar_compar_IET2017} tests different radar and sonar devices for detecting different classes of mobility via measuring micro-Doppler \cite{Sensing_Micro_Doppler_2006} sensitivity. In \cite{Sensing_radar_image_ICASSP2020}, a joint detection system that integrates a camera with an \ac{FMCW} radar is designed to realize object detection and 3D estimation. In \cite{Sensing_through_wall_RS2019}, a \ac{UWB} \ac{MIMO} radar equipped with manufactured Vivaldi antennas is designed and implemented to detect objects behind walls using \ac{SFCW} signals. Moreover, the variations of reflected signals can judge human motions even behind walls. The representative work is \cite{Sensing_Katabi_2013}, where a special method, termed \ac{ISAR}, is applied to deal with a moving object using a single receiver antenna. Recently, an IEEE group is working on an IEEE 802.11bf standard \cite{IEEE_802.11bf}, which aims to use existing \ac{WiFi} signals to realize sensing functions. There are commonly two methods for estimating human activities based on WiFi signals, namely \ac{RSSI} \cite{Sensing_RSSI_2014, Sensing_RSSI_JSTSP_2014, Sensing_RSSI_arXiv_2015, Sensing_RSSI_MobiHoc_2016} and \ac{CSI} \cite{WLAN_sensing_ACM_2019}. Although \ac{RSSI} has been successful in human activity detections, its coarse sensing resolution and high sensitivity to noise limit its applications in further areas. The second solution, \ac{CSI}, aims to extract amplitude \cite{WLAN_sensing_JIOT_2018} and phase information \cite{WLAN_sensing_PhaseU_Infocom2015, WLAN_sensing_PhaseBeat_ICDCS2017} to better assist human activity detections. In \cite{Sensing_WiFi_2017}, a WiFall system is designed to `see' human activities via measuring \ac{CSI}. A detailed propagation model is analytically studied to reveal the possibility of detecting human fall activities. In \cite{Sensing_WiFi_2016}, a WiHear system is designed to `hear' human talks based on micro-movement via radio reflections from mouth movements. In \cite{Sensing_WiFi_CSI_OFDM_MIMO_2018}, \ac{CSI} information is extracted from both \ac{OFDM} signals and \ac{MIMO} antennas. Therefore, detection accuracy is improved. In \cite{Sensing_WiFi_JIoT_2018}, \ac{CSI} from WiFi signals is extracted for monitoring vital signs and postures during sleeping.

It is noted that traditional radar signals are not initially designed for communications. Conversely, signalling for communications is not inherently designed to serve sensing functionalities. To achieve the joint sensing and communication purpose, communication radio signals and radar sensing signals have to be managed in \ac{TDM} mode, \ac{FDM} mode or \ac{SDM} mode. However, the multiplexing strategy will waste time, frequency or spatial resources. A number of approaches have emerged, aiming to design and test signalling that is appropriate for \ac{ISAC}. Work in \cite{RadarCom_INFOCOM_2020} proposed to use \ac{PSS} in the LTE frame for the radar sensing purpose. Work in \cite{RadarCom_RadarConf2017_partI} proposed a \ac{SDMA} scheme that can support radar and communications using the same transmit hardware with the same timing and spectral occupation. The principle behind the work is to send spatially orthogonal beams at the null space of the other one. Therefore, interference is avoided. This was further demonstrated experimentally in \cite{RadarCom_RadarConf2017_partII}, where analog-domain phased array antennas were employed to assist radar beam tracking and alignment. Work in \cite{RadarCom_access_2019} studied a new waveform design in \ac{ISAC}. The principle is to multiplex low out-of-band power emission signals with radar signals in frequency domain. However, this is a frequency multiplexing scheme and is not a dual-functional design. In addition, its experiment is based on single-antenna point to point links. Work in \cite{RadarCom_TGRS_2019} proposed to use mutually orthogonal waveforms via \ac{STC} in different beams for communication and radar rather than a single waveform beam. Work in \cite{RadarCom_TMTT_2019} aims to realize joint communication and radar functions in a \ac{FDM} mode via full-duplex in hardware based solutions. Work in \cite{RadarCom_OJVT_2021} designs a joint communication-radar experiment using single-carrier signals in a \ac{TDM} mode via full-duplex radar reception. Work in \cite{RadarCom_simulation_TSP_2020} proposed to achieve joint communication and radar functions by modulating information signalling onto standard radar waveforms through index modulation.

The main contribution of this work is to practically design and test over-the-air for the first time, a joint dual-functional radar communication waveform \cite{RadarCom_Fan_2018} for an integrated radar sensing and multi-user MIMO-OFDM communication system \cite{Tongyang_MU_MIMO_NB_IoT_2018}. Unlike existing work, the prototyping testbed in this paper can realize radar and communication using the same time, frequency and spatial resources. As a step ahead from \cite{RadarCom_Fan_2018}, the designed dual-functional \ac{ISAC} experiment in this work is based on the OFDM signal waveform, which enables a straightforward deployment of the \ac{ISAC} framework in many standard communication systems. Additionally, unlike pure theoretical simulations, this work obtains a practically working radar and communication trade-off factor that ensures radar beampattern quality and communication performance after comprehensive experiments on communication constellation diagrams, \ac{BER}, \ac{EVM} and radar beampattern quality.

The rest of this paper is organized as follows. Section \ref{sec:communication_model} will introduce the fundamentals of signal waveforms and multi-antenna communication architectures. In Section \ref{sec:ISAC_model}, the trade-off between radar sensing and communication is explained using the \ac{ISAC} model, followed by the radar beampattern illustrations in pure radar and pure communication systems. A multi-user MIMO-OFDM experiment is designed and implemented in Section \ref{sec:experiment_setup_validation} to verify the \ac{ISAC} framework in hardware. Finally, Section \ref{sec:conclusion} concludes the work.

\section{Communication Model} \label{sec:communication_model}

We consider a mutli-user MIMO-OFDM transmission, for which the received signal can be expressed as

\begin{equation}\label{eq:MIMO_tx_matrix_expression}
\mathbf{Y}=\mathbf{H}\tilde{\mathbf{X}}+\mathbf{W},
\end{equation}
where $\mathbf{H}=[\mathbf{h_{1}}, \mathbf{h_{2}},...,\mathbf{h_{N}}]\in\mathbb{C}^{K{\times}N}$ indicates a \ac{MIMO} channel matrix with $K$ being the number of receiver side users and $N$ being the number of transmitter side antennas. $\tilde{\mathbf{X}}=[\mathbf{x_{1}}, \mathbf{x_{2}},...,\mathbf{x_{L}}]\in\mathbb{C}^{N{\times}L}$ is the transmission symbol matrix after precoding, with $L$ being the number of time samples per data stream on each antenna. Similarly, the noise matrix $\mathbf{W}=[\mathbf{w_{1}}, \mathbf{w_{2}},...,\mathbf{w_{L}}]\in\mathbb{C}^{K{\times}L}$ indicates $K$ parallel noise vectors for $K$ receiver side users with $L$ noise samples per user.

The commonly used multicarrier signal format in 4G, 5G and \ac{WiFi} standards is \ac{OFDM}, which we employ in this work. Traditionally, each antenna is responsible for an OFDM symbol stream. Therefore, the symbol transmission matrix consists of $N$ parallel OFDM data streams with $L$ time samples for each data stream. The expression in \eqref{eq:MIMO_tx_matrix_expression} can be rewritten as

\begin{eqnarray}\label{eq:MIMO_tx_matrix_expression_rewritten_multicarrier}
\mathbf{Y}=\mathbf{X}+\underbrace{(\mathbf{H}\tilde{\mathbf{X}}-\mathbf{X})}_{MUI}+\mathbf{W},
\end{eqnarray}
where $\mathbf{X}\in\mathbb{C}^{K{\times}L}$ indicates the user side multicarrier symbol matrix. The second term in \eqref{eq:MIMO_tx_matrix_expression_rewritten_multicarrier} represents the \ac{MUI} term and the total power contributed by the MUI terms is computed as
\begin{eqnarray}\label{eq:P_MUI_multi_carrier}
{P}_{MUI}=\left\Vert \mathbf{H}\tilde{\mathbf{X}}-\mathbf{X}\right\Vert ^{2}_F,
\end{eqnarray}
where $\left\Vert\cdotp\right\Vert_F$ denotes the Frobenius matrix norm. The value of ${P}_{MUI}$ determines the value of \ac{SINR}. In order to have high throughput, the SINR should be maximized by minimizing the value of ${P}_{MUI}$.

An OFDM signal is expressed as 

\begin{equation}\label{eq:OFDM_discrete_signal}
X_k=\frac{1}{\sqrt{Q}}\sum_{m=1}^{M}s_{m}\exp\left(\frac{j2{\pi}mk}{Q}\right),
\end{equation}
where $X_k$ is the time sample with the index of $k=1,2,...,Q$, $M$ is the number of sub-carriers, $Q=\rho{M}$ indicates the number of time samples and $\rho$ is the oversampling factor. It is noted that $M{\leq}L$. $\frac{1}{\sqrt{Q}}$ is the normalization factor and $s_{m}$ is the $m^{th}$ single-carrier symbol in one OFDM symbol.

A matrix format can convert the expression in \eqref{eq:OFDM_discrete_signal} to the following
  
\begin{equation}\label{eq:tx_matrix_expression}
X=\mathbf{F}S,
\end{equation}
where $\mathbf{F}\in\mathbb{C}^{Q{\times}M}$ indicates a sub-carrier matrix with elements noted as $\exp(\frac{j2{\pi}mk}{Q})$ and $S\in\mathbb{C}^{M{\times}1}$ indicates the symbol vector with elements noted as $s_{m}$. The received signal, contaminated by \ac{AWGN}, $Z$, is expressed as
\begin{equation}\label{eq:tx_matrix_Channel_expression}
Y=\mathbf{F}S+Z,
\end{equation}
where $Y\in\mathbb{C}^{Q{\times}1}$ indicates one OFDM symbol. For an OFDM frame, we need to generate multiple OFDM symbols with overall $L$ time samples. In order to support a MIMO system defined in \eqref{eq:MIMO_tx_matrix_expression}, we need $K$ parallel OFDM signal generators. In this case, the user side symbol matrix ${\mathbf{X}}\in\mathbb{C}^{K{\times}L}$ is obtained. In the following, we will discuss the methodology of precoding ${\mathbf{X}}\in\mathbb{C}^{K{\times}L}$ to the dual-functional radar communication waveform $\tilde{\mathbf{X}}\in\mathbb{C}^{N{\times}L}$.

\begin{figure*}[ht]
\begin{center}
\includegraphics[scale=0.56]{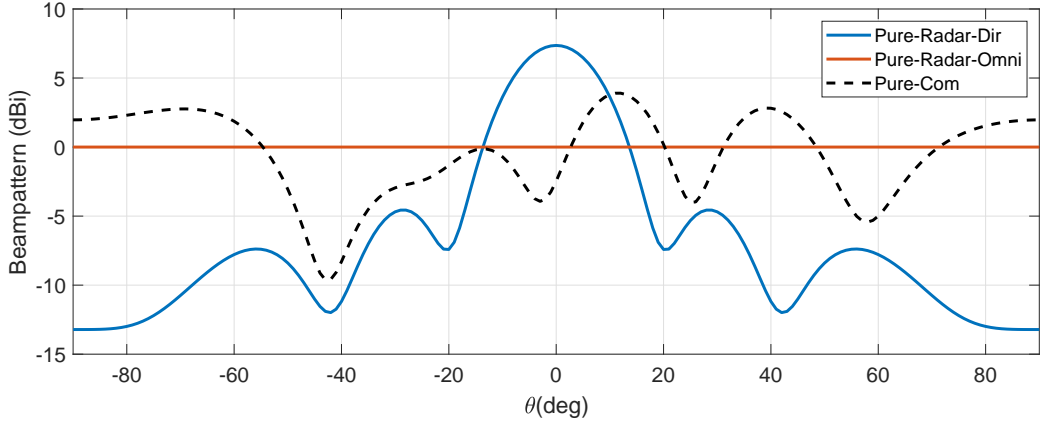}
\end{center}
\caption{Radar beampattern illustration for pure communication systems ($\gamma$=1) and pure radar systems ($\gamma$=0) considering \ac{OFDM} communication signals and directional/omnidirectional radar beampatterns. }
\label{Fig:RadarCom_simulation_beampattern_subs_overlapping}
\end{figure*}

\section{Trade-off Between Radar Sensing and Communications } \label{sec:ISAC_model}

To realize a dual-functional radar communication function, we employ the optimization methodology from \cite{RadarCom_Fan_2018} where a trade-off factor $\gamma$ is introduced to balance the performance of the communication part and the radar part. In this case, the resulting waveform can provide a balanced solution to both communications and radar waveform. 

We define the desired radar transmit signal as ${\mathbf{X}}_d$ where its design is detailed in \cite{RadarCom_Fuhrmann_2008}. The trade-off optimization problem considering the total power constraint is formulated as

\begin{equation}\label{eq:optimization_trade_off}
\begin{aligned}
\min_{\mathbf{\tilde{X}}} \quad &\gamma{\left\Vert \mathbf{H}\tilde{\mathbf{X}}-\mathbf{X}\right\Vert ^{2}_F}+(1-\gamma){\left\Vert \tilde{\mathbf{X}}-{\mathbf{X}}_d\right\Vert ^{2}_F}\\
\textrm{s.t.} \quad & \frac{1}{L}\left\Vert \tilde{\mathbf{X}}\right\Vert ^{2}_F =P_T,\\
\end{aligned}
\end{equation}
where the first term, $\left\Vert \mathbf{H}\tilde{\mathbf{X}}-\mathbf{X}\right\Vert ^{2}_F$ aims to minimize the MUI while the second term $\left\Vert \tilde{\mathbf{X}}-{\mathbf{X}}_d\right\Vert ^{2}_F$ aims to enforce the signal waveform to approach the desired radar waveform ${\mathbf{X}}_d$. ${0}\leq\gamma\leq{1}$ indicates the trade-off factor that balances the communication and radar performance. 

In general, there are two types of MIMO radar waveform designs. One is the orthogonal waveform, which generates omni-directional beampattern for searching unknown targets. Alternatively, MIMO radar may also track known targets via directional waveforms \cite{stoica2007probing}. Without loss of generality, in this paper we show that the proposed approach is capable of designing both orthogonal and directional MIMO radar waveforms while carrying communication information, which will be validated by experimental results.  

We can expand the two Frobenius norms and combine them in a single norm format as

\begin{equation}\label{eq:optimization_trade_off_expanding}
\begin{aligned}
\quad &\gamma{\left\Vert \mathbf{H}\tilde{\mathbf{X}}-\mathbf{X}\right\Vert ^{2}_F}+(1-\gamma){\left\Vert \tilde{\mathbf{X}}-{\mathbf{X}}_d\right\Vert ^{2}_F}\\
\quad & =\left\Vert [\sqrt{\gamma}\mathbf{H}^T, \sqrt{1-\gamma}\mathbf{I}_N]^T\tilde{\mathbf{X}}- [\sqrt{\gamma}\mathbf{X}^T, \sqrt{1-\gamma}\mathbf{X}^T_d]^T  \right\Vert ^{2}_F.\\
\end{aligned}
\end{equation}

To simplify the expression, we define $\mathbf{A}=[\sqrt{\gamma}\mathbf{H}^T, \sqrt{1-\gamma}\mathbf{I}_N]^T\in\mathbb{C}^{(K+N){\times}N}$, $\mathbf{B}=[\sqrt{\gamma}\mathbf{X}^T, \sqrt{1-\gamma}\mathbf{X}^T_d]^T\in\mathbb{C}^{(K+N){\times}L}$. Therefore, \eqref{eq:optimization_trade_off} can be reformulated as

\begin{equation}\label{eq:optimization_trade_off_simple}
\begin{aligned}
\min_{\mathbf{\tilde{X}}} \quad &\left\Vert \mathbf{A}\tilde{\mathbf{X}}-\mathbf{B}\right\Vert ^{2}_F\\
\textrm{s.t.} \quad & \left\Vert \tilde{\mathbf{X}}\right\Vert ^{2}_F =LP_T.\\
\end{aligned}
\end{equation}

While problem (\ref{eq:optimization_trade_off_simple}) is non-convex due to the quadratic equality constraint, it can be proved that strong duality holds, such that \eqref{eq:optimization_trade_off_simple} can be optimally solved via solving the dual problem \cite{RadarCom_Fan_2018}. To reduce the complexity incurred by the iterative algorithm of solving the dual problem, we consider a closed-form sub-optimal solution, which is obtained by using the simple least square method under the total power constraint as the following

\begin{equation}\label{eq:optimization_trade_off_final}
\tilde{\mathbf{X}}= \frac{\sqrt{LP_T}}{\left\Vert \mathbf{A}^{\dag}\mathbf{B}\right\Vert_F}\mathbf{A}^{\dag}\mathbf{B},
\end{equation}
where $\left(\cdot\right)^\dag$ represents the pseudo inverse of the matrix. To illustrate the trade-off performance for the omnidirectional beampattern and directional beampattern designs, we will use `Pure-Radar-Omni' and `Pure-Radar-Dir' to represent pure omnidirectional and directional radar beampattern, respectively. We will use `Pure-Com' to represent the radar beampattern when pure communication is enabled. For dual-functional radar communication systems, we will use terms `RadarCom-Omni' and `RadarCom-Dir' correspondingly.

\begin{figure*}[ht]
\begin{center}
\includegraphics[scale=0.47]{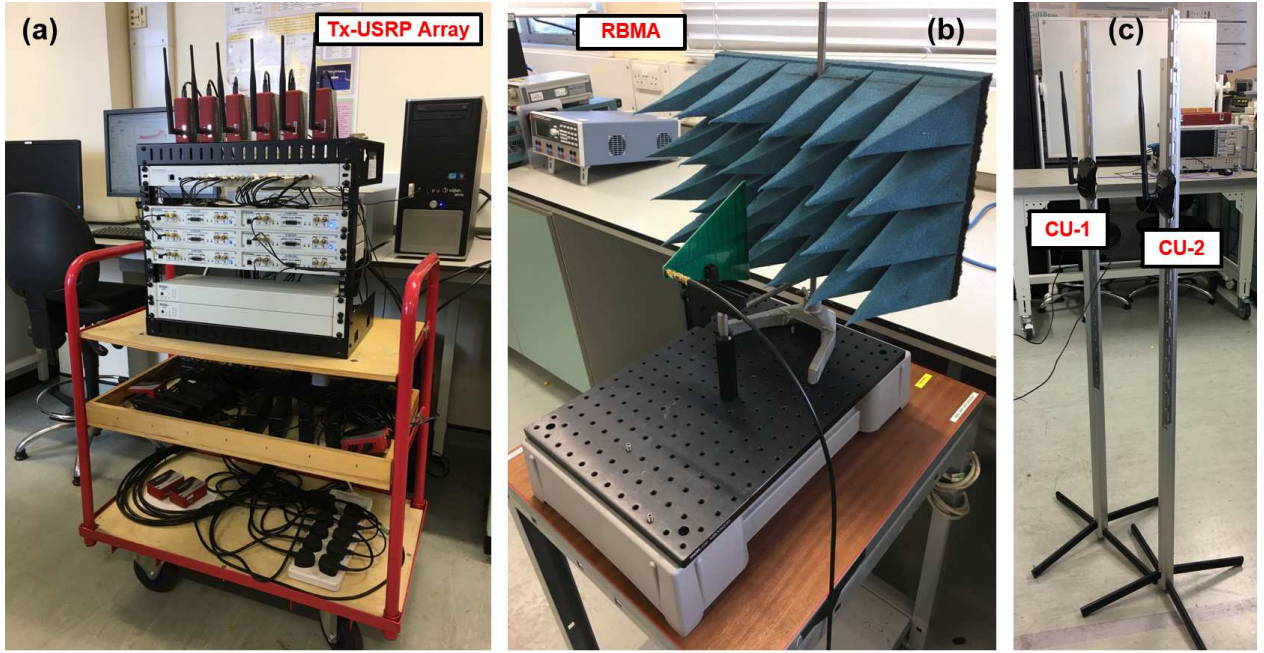}
\end{center}
\caption{Experiment platform setup. (a) Tx-USRP Array: MIMO transceiver that precodes and decodes multi-user signals. (b) Radar Beampattern Measurement Apparatus (RBMA): a directional antenna to measure radar beampattern. (c) CU-1 and CU-2: two omnidirectional antennas to receive communication signals. }
\label{Fig:Experiment-floor-plan-I}
\end{figure*}

The trade-off performance for pure communication systems ($\gamma=1$) and pure radar systems ($\gamma=0$) are demonstrated in Fig. \ref{Fig:RadarCom_simulation_beampattern_subs_overlapping}. It is obvious from \eqref{eq:optimization_trade_off} that when the trade-off factor $\gamma=0$, the intended waveform will match closely the perfect radar waveform as shown in Fig. \ref{Fig:RadarCom_simulation_beampattern_subs_overlapping} while it will be far away from the communication featured waveform. In this case, the scenarios with $\gamma=0$ would cause performance degradation in communications. When the trade-off factor is increased to $\gamma=1$, the radar part in \eqref{eq:optimization_trade_off} will be removed. Therefore, the communication part dominates the integrated system and the intended waveform will be more likely to follow the optimal communication constraints. In this case, $\gamma=1$ leads to pure communication scenarios and Fig. \ref{Fig:RadarCom_simulation_beampattern_subs_overlapping} reveals that the `Pure-Com' radar beampattern is more likely to be random, which is far away from `Pure-Radar-Omni' and `Pure-Radar-Dir' radar beampatterns. 

For other values of $\gamma$, trade-off exists between communication and radar performance. Explicitly, as $\gamma$ is increased, priority is given to communications at the expense of radar performance, and vice versa. Fig. \ref{Fig:RadarCom_simulation_beampattern_subs_overlapping} merely shows the general design principle. The variations of communication BER and radar beampattern at different values of $\gamma$ will be investigated using our experiment testbed in the following sections.

\begin{figure*}[ht]
\begin{center}
\includegraphics[scale=0.52]{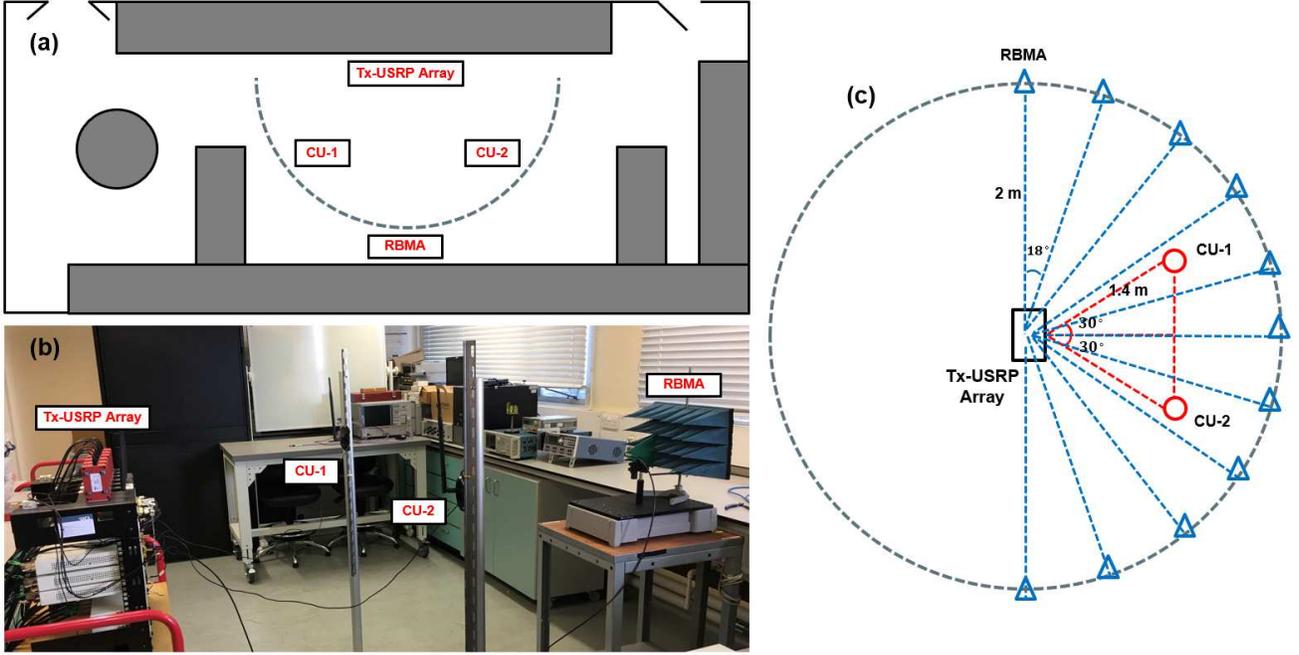}
\end{center}
\caption{Experiment measurement setup. (a) Laboratory floor plan. (b) Platform units arrangement. (c) Radar beampattern measurement. }
\label{Fig:Experiment-floor-plan-II}
\end{figure*}

\section{Experiment Setup and Validation} \label{sec:experiment_setup_validation}

\subsection{Experiment Platform Setup} \label{subsec:experiment_platform_setup}

As demonstrated in Fig. \ref{Fig:Experiment-floor-plan-I}, the designed platform is a $6\times{2}$ MIMO-OFDM system working at 2.4 GHz carrier frequency, consisting of a Tx-USRP array (USRP cluster with antenna array), two communication users (two antennas associated with two separate USRPs), a radar beampattern measurement apparatus (a radar beampattern measurement detector associated with a stand-alone USRP).

\subsubsection{Tx-USRP Array}

The emulated base station, noted as the Tx-USRP array, consists of six USRP-RIO-2953R. Each of the devices has two \ac{RF} chains, in which one can be used for signal generation and the other one is for signal reception. In this experiment, we use one RF chain from each USRP for signal generation at the carrier frequency $f_{RF}$=2.4 GHz with the sampling rate of 20 MS/s. The symbol modulated at each sub-carrier is QPSK. The number of data sub-carriers is 76 and the \ac{IFFT} size is 128. In addition, each OFDM symbol also considers 10 \ac{CP} samples for the mitigation of channel effects. The output from each USRP is fed to an omnidirectional antenna via a Vaunix LPS-402 programmable phase shifter \cite{Vaunix_phase_shifter_ref}. In this experiment, the phase shifter is merely used for holding the omnidirectional antenna without any phase control functions. However, the activation of the phase control function in each phase shifter will enable a more power efficient hybrid analog-digital multi-user MIMO system design \cite{Tongyang_hybrid_precoding_2020}, which could be the future research direction of \ac{ISAC}. In total, six antennas are placed in a \ac{ULA} format at the top of the testbed with the spacing of half wavelength. As mentioned, the second RF chain in each USRP can be reserved for signal reception. Therefore, the experiment platform can support up to a $6\times{6}$ MIMO-OFDM system.

\subsubsection{Communication Users (CUs)}

In this experiment, for simplicity we emulate two users as the receiver side. Therefore, the communication part is a $6\times{2}$ MIMO-OFDM system, which includes the Tx-USRP array base station and two CUs at the receiver side. Each CU, equipped with an omnidirectional antenna, is connected to a USRP, which can separately process received signals.

\subsubsection{Radar Beampattern Measurement Apparatus (RBMA)}

Due to restricted measurement environment, the entire experiment has to be managed in an indoor laboratory. Therefore, the radar beampattern measurement will be affected due to signal reflections in such a small indoor space. Nevertheless, measuring the beampattern in an anechoic chamber would significantly hamper our ability to investigate the multipath effects in the performance of the CUs. Therefore, we carried out the experiment in the indoor laboratory to allow rich multipath propagation. For the radar part, to maximally mitigate the multipath effect, obtain an accurate measurement of the LoS power, and emperically measure the transmitted beampattern, we employ a directional 6 dBi log-periodic (LP) PCB antenna, which works from 850 MHz to 6.5 GHz carrier frequency. To further reduce reflections that would cause pattern measurement errors, we also place radiation absorbing material (RAM) behind the LP antenna. The RAM can efficiently reduce signal reflections with central frequency from 1 GHz to 40 GHz.

\begin{figure*}[ht]
\begin{center}
\includegraphics[scale=0.3]{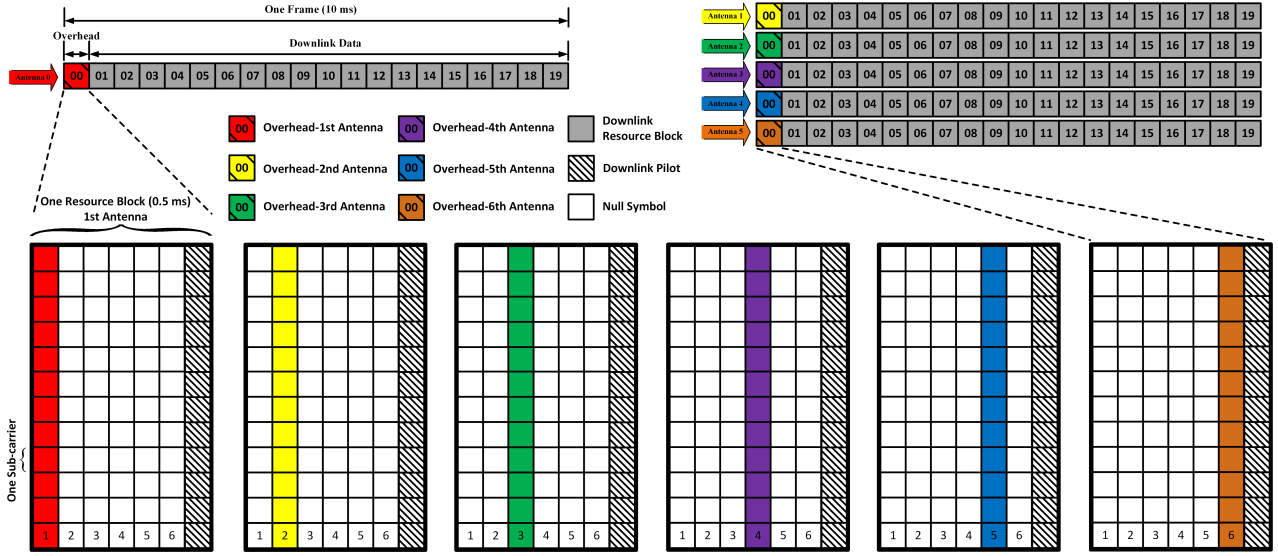}
\end{center}
\caption{Frame and resource block structure for the dual-functional radar and communication multiuser MIMO system.}
\label{Fig:frame_RB_structure}
\end{figure*}

\subsection{Experiment Measurement Setup}\label{subsec:experiment_measurement_setup}

The floor plan of the experiment setup is shown in Fig. \ref{Fig:Experiment-floor-plan-II}(a), where the indoor laboratory is approximately 4 m wide and 9 m long. There are plenty of objects that could cause signal reflection and blocking. For the communication part, this is not an issue since wireless channel will be measured in real time and precoding will be applied correspondingly to mitigate the MIMO transmission interference. Moreover, this experiment employs OFDM signals instead of single-carrier signals. Therefore, the multicarrier structure of OFDM signals is robust to multipath channel effect. Commonly, pilot symbols are used to estimate \ac{CSI}, which will be used to equalize channel effects using a one-tap equalizer. 

The three-dimensional platform setup for the communication part is the following. The location for the Tx-USRP Array and CUs are labelled in Fig. \ref{Fig:Experiment-floor-plan-II}(b) where two users are placed in front of the base station. The horizontal distance between two users is 1.4 m and each user is 1.4 m away from the base station. Due to the limited space, to obtain sufficient measurement at various \ac{SNR}, we will maintain the total signal transmission power while adjusting noise for each measurement. In the experiment, SNR from each antenna is different. Therefore, we measure received SNR at the user side. The antenna array at the base station is 1.5 m above the floor and the two users are placed 1.5 m above the floor as well. For the radar beampattern antenna, it is placed 2 m away from the base station and 1.5 m above the floor.

The radar beampattern measurement strategy is demonstrated in Fig. \ref{Fig:Experiment-floor-plan-II}(c) in which the radar beampattern measurement apparatus will measure the signal power every 18 degrees with a radius of 2 m. Therefore, it will measure 10 points considering 180 degrees. To ensure accurate measurement, the radar beampattern measurement apparatus is placed at the same height with the base station antenna array. It should be noted that the experiment platform employs small dipole antennas for signal transmission. The size of each dipole antenna is similar to one wavelength $\lambda$ when considering the carrier frequency $f_{RF}$=2.4 GHz. Therefore, the beampattern measurement at a 2 m distance is greater than $2\lambda$ \cite{book_antenna-theory-analysis-4th} and is practically within the far-field range. In addition, to mitigate multipath effect to the beampattern measurement, we use a 6 dBi LP PCB antenna, which has a narrow and focused radiation beam. Therefore, it can focus on the \ac{LOS} signal collection within its beam range and avoid potential multipath signal collections from other reflection directions. For the radar part, we have a stand-alone USRP for radar beampattern power measurement.

The radar beampattern power computation is based on received symbols. Since each symbol has real and imaginary parts, therefore the power will be calculated as  
\begin{equation}\label{eq:rx_beam_power_computation}
P=\frac{1}{{Q}}\sum_{k=1}^{Q}[\Re(X(k))^2+\Im(X(k))^2], 
\end{equation}
where $X(k)$ indicates the $k^{th}$ received complex symbol and $\frac{1}{\sqrt{Q}}$ is the scaling factor for average power computation. $\Re{\cdotp}$ and $\Im{\cdotp}$ indicate the real part and imaginary part of a symbol, respectively.

\subsection{Frame Design and Channel Estimation}

\subsubsection{Frame Structure}

The frame structure for this experiment follows the same structure in \cite{Tongyang_MU_MIMO_NB_IoT_2018} and is shown in Fig. \ref{Fig:frame_RB_structure} where 20 resource blocks are combined to form one frame. The time duration for one frame is 10 ms. The first resource block is reserved for signalling overhead, which will be used mainly for MIMO channel estimation. Each resource block includes seven OFDM symbols and the first resource block has a unique OFDM symbol allocation scheme. The interference from MIMO antennas can be solved via transmitter precoding based on estimated CSI, which indicates the importance of accurate CSI estimation. To avoid interference to CSI estimation, we multiplex the overhead at each antenna in time-domain as illustrated in Fig. \ref{Fig:frame_RB_structure}. In this case, even though the data part is interfered, the overhead part is interference free. It should be noted that to mitigate potential channel and hardware imperfections, an additional downlink pilot is applied for all the data streams.

\begin{figure*}[ht]
\begin{center}
\includegraphics[scale=0.5]{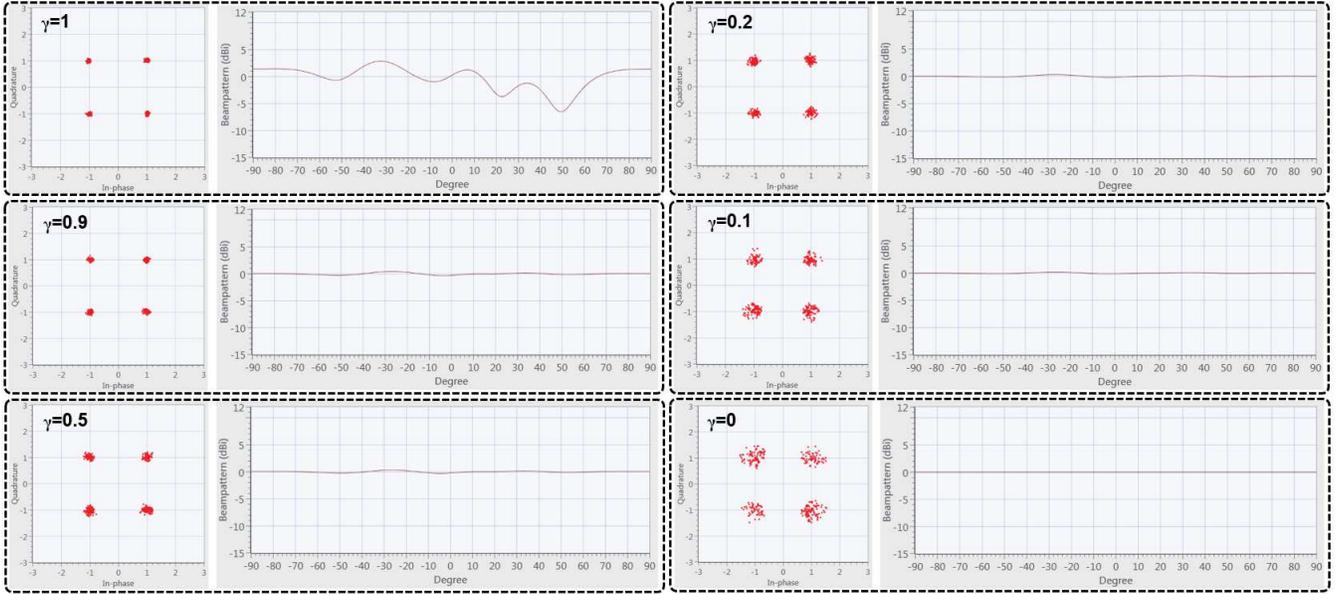}
\end{center}
\caption{Experiment results in user side constellation diagram measurements and Tx-USRP side radar omnidirectional beampatterns. }
\label{Fig:Experiment-USRP-constellation_beam_OmniDir}
\end{figure*}

\subsubsection{MIMO Channel Estimation}

Based on the interleaved overhead structure in Fig. \ref{Fig:frame_RB_structure}, we define a pilot matrix as \begin{equation}
\mathbf{P} =
 \begin{bmatrix}
  p_{1} & 0  & 0& 0& 0& 0\\
  0     & p_{2} & 0& 0& 0& 0\\
  0     & 0 & p_{3} & 0& 0& 0\\
  0     & 0& 0 & p_{4} & 0& 0\\
  0     & 0 & 0& 0& p_{5} & 0\\
  0     & 0& 0& 0 & 0 & p_{6} \\
 \end{bmatrix},
\label{eq:X_matrix}\end{equation}
where $p_1,p_2,p_3,p_4,p_5,p_6$ are one pilot symbol at each antenna.

The system is in a $2\times{6}$ channel model with the matrix format as
\begin{equation}
\mathbf{H} =
 \begin{bmatrix}
  h_{11} & h_{12} & h_{13} & h_{14}& h_{15} & h_{16}\\
  h_{21} & h_{22} & h_{23} & h_{24}& h_{25} & h_{26}
 \end{bmatrix}.
\label{eq:H_matrix}\end{equation}

Therefore, after the MIMO channel and \ac{AWGN} contamination, the received symbol matrix is expressed as

\begin{equation}
 \begin{bmatrix}
  y_{11} & y_{12}&y_{13} & y_{14}&y_{15} & y_{16}   \\
  y_{21} & y_{22}&y_{23} & y_{24}&y_{25} & y_{26}   \\
 \end{bmatrix}=\mathbf{H}\mathbf{P}+
 \begin{bmatrix}
  z_{11} & z_{21}\\
  z_{12} & z_{22}\\
  z_{13} & z_{23}\\
  z_{14} & z_{24}\\
  z_{15} & z_{25}\\
  z_{16} & z_{26}
 \end{bmatrix}^{T},
\label{eq:Y_matrix}\end{equation} 
where $y_{m,n}$ indicates the received symbols at the $m^{th}$ user from the $n^{th}$ antenna. Therefore, the MIMO channel matrix can be calculated regardless of noise via

\begin{equation}
\mathbf{\hat{H}} =
 \begin{bmatrix}
  y_{11}/p_{1} & y_{12}/p_{2} &  y_{13}/p_{3} & y_{14}/p_{4}&  y_{15}/p_{5} & y_{16}/p_{6}\\
  y_{21}/p_{1} & y_{22}/p_{2} &  y_{23}/p_{3} & y_{24}/p_{4}&  y_{25}/p_{5} & y_{26}/p_{6}
 \end{bmatrix}.
\label{eq:H_estimate_matrix}\end{equation} 

Based on the estimated MIMO channel matrix $\mathbf{\hat{H}}$, the Tx-USRP Array can do signal precoding such that MIMO antenna interference can be avoided.

\subsubsection{OFDM Channel Estimation}

Multicarrier signals can effectively convert a multipath effect into an equivalent diagonal composite matrix due to the use of Fourier transform. Unlike the MIMO channel matrix $\mathbf{{H}}$, we define a new channel matrix $\mathbf{{G}}$ for the multipath effect. Following the \ac{AWGN} channel distorted signal expression in \eqref{eq:tx_matrix_Channel_expression}, the new expression considering multipath is give by

\begin{equation}\label{eq:tx_matrix_G_Channel_expression}
Y=\mathbf{G}\mathbf{F}p_{g}+Z,
\end{equation}
where $p_{g}$ is the downlink pilot symbol defined in Fig. \ref{Fig:frame_RB_structure}. 

After demodulation at the receiver, the distorted signal is expressed as

\begin{equation}\label{eq:rx_G_matrix_expression}
R_g=\mathbf{F^*}\mathbf{G}\mathbf{F}p_{g}+\mathbf{F^*}Z=\mathbf{D}p_{g}+W,
\end{equation}
where $\mathbf{D}$=$\mathbf{F^*}\mathbf{G}\mathbf{F}$ is an diagonal matrix, in which its diagonal elements $diag(\mathbf{D})$ can be extracted for one-tap equalization such that multipath, imperfect timing, phase offset and power distortion will be removed.

\subsection{Experiment Results}

\begin{figure*}[ht]
\begin{center}
\includegraphics[scale=0.5]{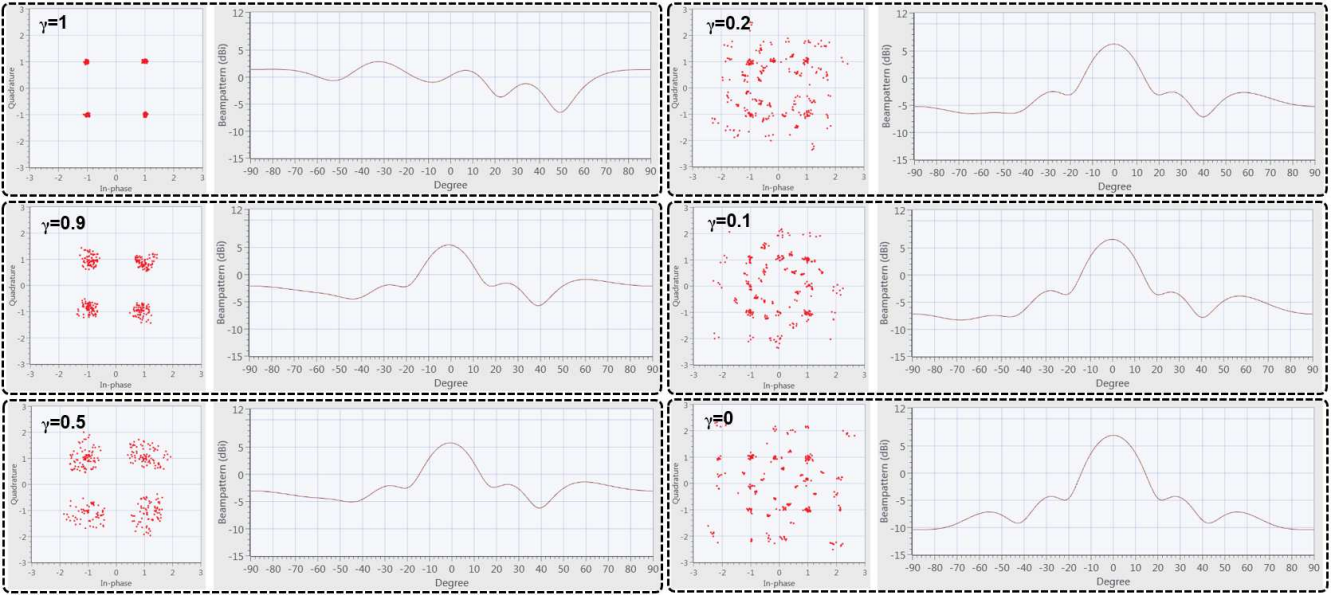}
\end{center}
\caption{Experiment results in user side constellation diagram measurements and Tx-USRP side radar directional beampatterns. }
\label{Fig:Experiment-USRP-constellation_beam_Dir}
\end{figure*}

We practically verify the trade-off between communications and radar functions. The Tx-USRP Array, two receiver side users and the radar beampattern measurement apparatus are placed following the floor plan in Fig. \ref{Fig:Experiment-floor-plan-II}(a). At the beginning, we will not add \ac{AWGN} to the testbed and just check the functionality of the communication part and radar part. To evaluate the trade-off between communication and radar beampattern, we test different representative values for $\gamma$ such as $\gamma$=1, 0.9, 0.5, 0.2, 0.1, 0.

\begin{figure*}[ht]
\begin{center}
\includegraphics[scale=0.57]{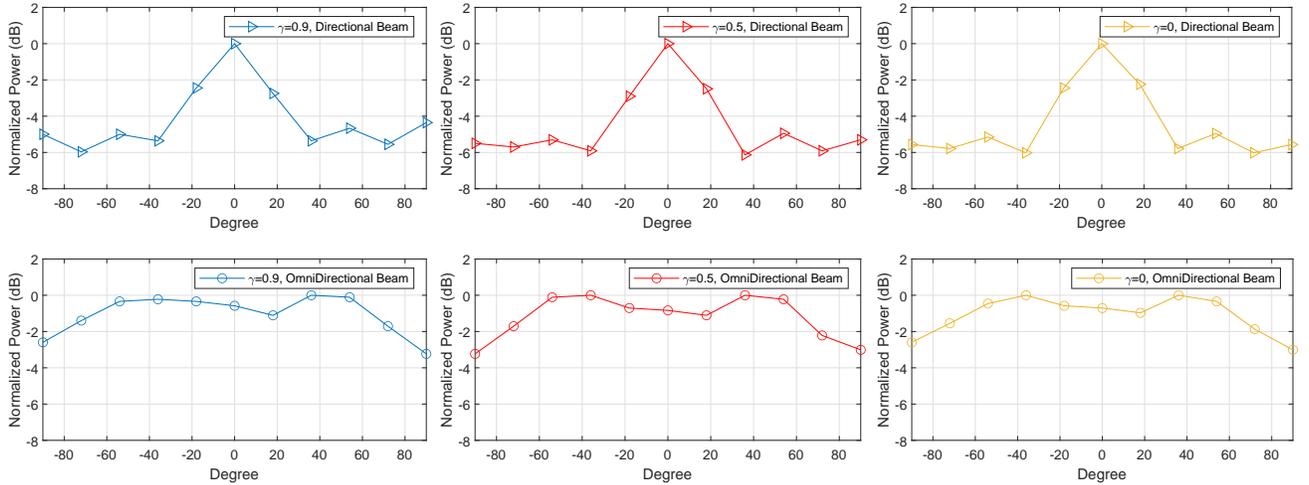}
\end{center}
\caption{Measured Omnidirectional and directional radar beampatterns at the radar beampattern measurement apparatus when $\gamma$=0.9, 0.5, 0. }
\label{Fig:RadarCom_experiment_USRP_beampattern_subplot}
\end{figure*}

The measured results for omnidirectional systems are shown in Fig. \ref{Fig:Experiment-USRP-constellation_beam_OmniDir} where different values of $\gamma$ are configured. In this result, we plot the theoretical transmitter side beampattern from Tx-USRP Array based on the estimated MIMO channel matrix in \eqref{eq:H_estimate_matrix}. Since the measurement at this point is not affected by over-the-air multipath, its beampattern is relatively ideal, which can be used as a benchmark for practical beampattern measurement. It is observed that small values of $\gamma$ will degrade communication performance evidenced by the fact that the constellation points are more scattered. However, radar beampattern becomes better when the value of $\gamma$ reduces. With the increase of $\gamma$, constellation points are more focused leading to potentially better BER performance but at the cost of more distorted radar beampatterns. Accordingly, it is clear that there is a trade-off configuration between communication performance and radar beampattern quality. Based on the observation in Fig. \ref{Fig:Experiment-USRP-constellation_beam_OmniDir}, it is obvious that when $\gamma$ is large sufficient, it will provide a pure radar beampattern achievable waveform. The optimal trade-off value can be reduced to $\gamma$=0.9 such that constellation points are clearly separated while the radar beampattern is still roughly in a perfect beampattern shape.

For the directional system measurement illustrated in Fig. \ref{Fig:Experiment-USRP-constellation_beam_Dir}, its trade-off variation is more obvious. When $\gamma$=0.2, 0.1, 0, the communication is greatly degraded since constellation points are scattered and rotated. However, the above systems show high quality radar beampattern. When $\gamma$ is increased to 0.5, constellation points start to appear. Further increasing $\gamma$ to 0.9, the constellation becomes even better with a reasonable radar beampattern. Therefore, similar to the omnidirectional results in Fig. \ref{Fig:Experiment-USRP-constellation_beam_OmniDir}, $\gamma$=0.9 is an optimal setup for directional radar where still a clear beam is obtained with a peak to side lobe ratio (PSLR) of 6 dB. It is noted that $\gamma$=1 indicates pure communication systems. Therefore, the obtained beampatterns for the directional and omnidirectional design in Fig. \ref{Fig:Experiment-USRP-constellation_beam_OmniDir} and Fig. \ref{Fig:Experiment-USRP-constellation_beam_Dir} are identical.

Beampatterns measured by the radar beampattern measurement apparatus are also included in Fig. \ref{Fig:RadarCom_experiment_USRP_beampattern_subplot} where it is obvious that three directional beampattern in Fig. \ref{Fig:RadarCom_experiment_USRP_beampattern_subplot} show pattern peaks at $\theta$=$0^{\degree}$ while all other angles have lower normalized power. In this case, a directional beampattern is obtained and it is in good agreement with the ideal pattern at Tx-USRP Array in Fig. \ref{Fig:Experiment-USRP-constellation_beam_Dir}, but with a reduced PSLR of about 4 dB. For the omnidirectional beampattern measured in Fig. \ref{Fig:RadarCom_experiment_USRP_beampattern_subplot}, the measured patterns are not flat as expected from Fig. \ref{Fig:Experiment-USRP-constellation_beam_OmniDir} but are within 3 dB.

\begin{figure}[t!]
\begin{center}
\includegraphics[scale=0.58]{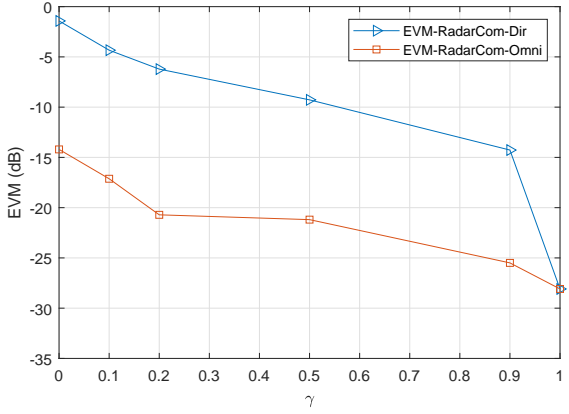}
\end{center}
\caption{Measured EVM versus $\gamma$ for omnidirectional and directional RadarCom systems. }
\label{Fig:RadarCom_experiment_USRP_EVM}
\end{figure}

Constellation diagram is a visual way to tell the performance of communications. Fig. \ref{Fig:RadarCom_experiment_USRP_EVM} compares constellation performance in terms of \ac{EVM} for omnidirectional and directional RadarCom systems under different values of $\gamma$. As expected, the EVM for both omnidirectional and directional systems becomes better with the increase of $\gamma$. In addition, the EVM of omnidirectional systems outperforms directional systems at all values of $\gamma$. Moreover, it is observed that the variations of $\gamma$ have smooth impact on omnidirectional communication systems while its value changing has great effect on directional communication systems especially when the value of $\gamma$ is between 0.9 and 1. The above discoveries can be explained based on the observations from Fig. \ref{Fig:Experiment-USRP-constellation_beam_OmniDir} and Fig. \ref{Fig:Experiment-USRP-constellation_beam_Dir}.

\begin{figure}[t!]
\begin{center}
\includegraphics[scale=0.6]{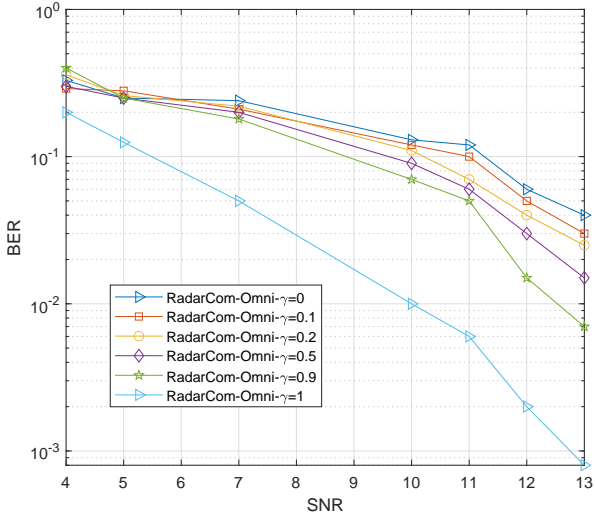}
\end{center}
\caption{Experiment BER measurement for different omnidirectional systems. }
\label{Fig:RadarCom_experiment_USRP_BER_omni}
\end{figure}

\begin{figure}[t!]
\begin{center}
\includegraphics[scale=0.6]{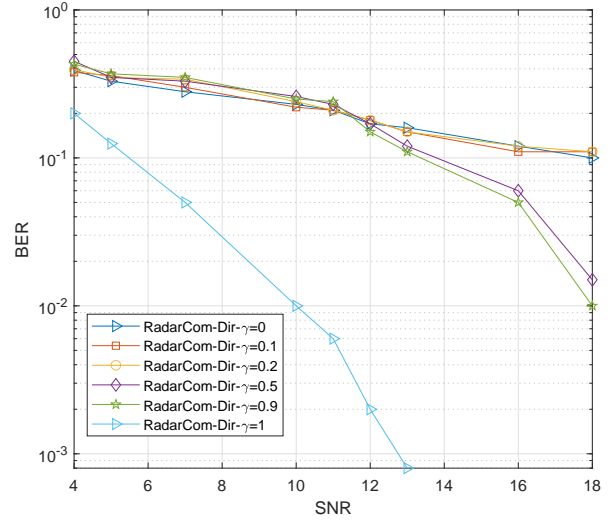}
\end{center}
\caption{Experiment BER measurement for different directional systems. }
\label{Fig:RadarCom_experiment_USRP_BER_dir}
\end{figure}

For a complete BER versus SNR measurement, we maintain the total transmission signal power and tune noise power such that various SNRs can be obtained. Fig. \ref{Fig:RadarCom_experiment_USRP_BER_omni} demonstrates the measured BER for omnidirectional systems. It is clear that the pure radar scenario shows the worst BER performance when $\gamma$=0. The pure communication scenario shows the best performance when $\gamma$=1. All other values of $0<\gamma<1$ show BER performance in the middle. As explained before, the optimal value of $\gamma$ is 0.9, where it is approximately 2-3 dB gap with the pure communication system when $\gamma$=1. Jointly considering the measured beampattern of $\gamma$=0.9 in Fig. \ref{Fig:RadarCom_experiment_USRP_beampattern_subplot}, it is inferred that a RadarCom system can realize dual functional communications and radar sensing at the cost of BER performance loss. BER variations for directional systems are more obvious in Fig. \ref{Fig:RadarCom_experiment_USRP_BER_dir} where the performance gap between $\gamma$=1 and $\gamma$=0.9 is increased to around 8 dB. The reason can be observed by constellations in Fig. \ref{Fig:Experiment-USRP-constellation_beam_Dir} and EVM in Fig. \ref{Fig:RadarCom_experiment_USRP_EVM}

\section{Conclusion} \label{sec:conclusion}

This work designed an over-the-air multi-user MIMO-OFDM testbed to validate a dual-functional radar sensing and communication waveform. Over-the-air experiments reveal that an optimal trade-off factor is available to balance the performance for both radar and communication functions. Practical measured radar beampatterns have reasonable radiation shape compared to simulation results. The BER results demonstrate a minor performance loss relative to pure communications when omnidirectional radiation waveform is applied while the performance loss is widened when directional radiation waveform is used.

\bibliographystyle{IEEEtran}
\bibliography{paper_Ref}

\end{document}